\documentstyle[12pt]{article}
\begin{document}


\renewcommand{\thesection}{\arabic{section}}
\renewcommand{\thesubsection}{\thesection.\arabic{subsection}}
\renewcommand{\theequation}{\arabic{equation}}
\renewcommand {\c}  {\'{c}}
\newcommand {\cc} {\v{c}}
\newcommand {\s}  {\v{s}}
\newcommand {\CC} {\v{C}}
\newcommand {\C}  {\'{C}}
\newcommand {\Z}  {\v{Z}}

\baselineskip=24pt


\begin{center}
{\bf  Permutation Invariant Algebras, a Fock Space Realization and the Calogero 
Model}
 
\bigskip

S.Meljanac\footnote{e-mail: meljanac@thphys.irb.hr}, 
 M.Milekovi\'c\footnote{e-mail: marijan@phy.hr }
 and M.Stoji\'c \footnote{e-mail: mstojic@thphys.irb.hr}\\

$^{1,3}$ Rudjer Bo\v{s}kovi\'c Institute, Bijeni\v cka  c.54, 10002 Zagreb,
Croatia\\[3mm] 

$^{2}$ Prirodoslovno-Matemati\v{c}ki Fakultet, Fizi\v{c}ki Zavod, P.O.B. 331,
 Bijeni\v{c}ka c.32,
\\ 10002 Zagreb, Croatia \\[3mm]

\bigskip

Short Title:  Permutation Invariant Algebras and the Calogero 
Model

\end{center}
\setcounter{page}{1}
\bigskip

{\bf Abstract.} We study permutation invariant oscillator algebras and their 
Fock space representations using three equivalent techniques, i.e. 
(i) a normally ordered expansion in creation and annihilation operators, (ii) the action  
of annihilation operators on monomial states in Fock space and (iii)  Gram matrices of inner products
in Fock space. We  separately discuss permutation invariant algebras which possess hermitean
number operators and permutation invariant algebras which possess non-hermitean number operators. The
results of a general analysis are applied to the $S_M$-extended Heisenberg algebra, underlying the $M$-body Calogero 
model. Particular attention is devoted to the analysis of Gram matrices for the Calogero model. We discuss 
their structure, eigenvalues and eigenstates. We obtain a general condition for positivity of eigenvalues, 
meaning that all norms of states in Fock space are positive if this condition is satisfied. We find
a universal critical point at which the reduction of the physical degrees of freedom occurs. We construct dual 
operators, leading to the ordinary Heisenberg algebra of free Bose oscillators.  From 
the Fock-space point of view, we briefly discuss the existence of mapping from the Calogero oscillators to the free 
Bose oscillators and vice versa.
 

\newpage



\section {Introduction}

The classical and quantum integrable model of $M$ interacting particles on a
line, introduced by Calogero [1], has been intensively studied  during the   past
 few years. The model and its generalizations [2] are connected with a number of
physical problems, ranging from condensed matter physics [3] to gravity and 
  black-hole physics [4]. The algebraic structure of the Calogero model and its
successors, studied earlier using group theoretical methods [5], has recently been 
 reconsidered by a number of authors in the framework of the $S_M$ -
extended Heisenberg algebra [6].\\
Apart from its particular realization,  the $S_M$ - extended Heisenberg algebra is 
basically ${\it {a}}$  ${\it{multi-mode}}$  ${\it{oscillator} }$ algebra with 
${\it {permutation}}$ 
${\it {invariance} }$. The general techniques for analysing such  a class of oscillator 
algebras were developed earlier in a series of papers [7,8].\\
In the  present paper we apply these techniques to the Calogero model in its
operator formulation. We start our analysis with the two-body Calogero model. 
On the algebraic grounds, this model is described by a particular class of  deformed
single-mode oscillator algebras which were treated in a unified manner in [9].\\
In Section 2, we describe a single-mode algebra underlying the two-body Calogero model. 
We construct a mapping from this algebra to the ordinary Bose algebra. We also present 
the number operator $N$ and the exchange operator $K$ as ( infinite ) series in creation and 
annihilation operators. At the end of Section 2 we construct an algebra that is dual to 
the original algebra of the two-body Calogero model.\\
In Section 3 we discuss general multi-mode oscillator algebras with permutation
invariance. We describe  two distinct classes of these algebras: (i) algebras which possess 
well-defined ${\it hermitean}$ number operators ( i.e. transition number operators $N_{ij}$,
partial number operators $N_{ii}\equiv N_i$ and the total number operator $N\equiv \sum N_i$ ),
 and (ii) algebras which   possess
well-defined number operators  but for which only the total number operator $N$ is hermitean.
The analysis of these algebras is performed in three equivalent ways: using  (i)  a normally
ordered expansion in creation and annihilation operators, 
(ii) the  action of annihilation operators on  monomial states in Fock space, and 
(iii)  Gram matrices of scalar products in Fock space . 
We conclude  Section 3 with a discussion of the  general structure of 
transition  number operators $N_{ij}$ and exchange operators $K_{ij}$. \\
The ideas developed in the preceding Sections are applied to the many-body Calogero model in 
Section 4. The algebra underlying the many-body Calogero model ( $S_M$-extended Heisenberg
algebra ) is discussed along the lines described  in Section 3. Special attention is devoted
to the analysis of Gram matrices and to the construction  of number operators
and exchange operators as an infinite series in creation and annihilation operators.
We also find that the $S_M$-extended Heisenberg algebra can be defined as 
a generalized triple operator algebra. Generalizing the construction given in Section 2 to the
multi-mode case, we define and investigate the structure of dual algebras.  Section 4 
ends with a short discussion of mappings from the Calogero algebra 
to the set of free bosonic oscillators.
Section 5 is a short summary.




 \section { Two-body Calogero model and  deformed \\ single-mode oscillator algebras }
 Particular aspects of  single-mode deformed oscillator algebras were 
 studied by  a number of authors, starting with seminal papers 
 [10].
  A  unified view of deformed single-mode oscillator algebras 
 was   proposed in [9,11].\\
 Basically, these algebras are generated by a set of generators $\cal {G}$, 
 involving  annihilation ($a$) and  creation ($a^{\dagger}$) operators,
  together with the well-defined number operator $N$:
$$
{\cal {G}}:=   \{ 1 , a , a^{\dagger}, N \},\\[4mm]
$$
$$( a )^{\dagger} = a^{\dagger} \qquad N=N^{\dagger},
$$
The following commutation relations hold:
\begin{equation}
\begin{array}{c}
[N,a] = -a,\qquad [N,a^{\dagger}] = a^{\dagger},\\[4mm]
[N,a^{\dagger}a] = [N,aa^{\dagger}]=0,\\[4mm]
 a a^{\dagger} - q a^{\dagger} a = G(N),
\end{array}
\end{equation}
where $q \in \bf {R}$ and $G(N)$ is the hermitean, analytic function of the number
 operator.  
Vacuum conditions are $a|0\rangle =0$ and $N|0\rangle =0$, with $\langle 0|0 \rangle =1$.
Since $[N,a^{\dagger}a] = [N,aa^{\dagger}]=0$, we can write
\begin{equation}
\begin{array}{c}
a^{\dagger}a=\phi (N),\\[4mm]
aa^{\dagger}=\phi (N+1) ,
\end{array}
\end{equation}
 where $ \phi(N) \geq 0$ is some function of the number operator. Several examples of algebras that
 belong to the class  (1) 
 and their corresponding functions $ \phi(N)$ are given in [9].
 
 Here we want to discuss a variant of the algebra (1) ,
 with $G(N)=1+2\nu K$ and $q=1$,  namely
 \begin{equation}
\begin{array}{c}
a a^{\dagger} -  a^{\dagger} a =  1 + 2 \nu K , \qquad \nu \in {\bf R}\\[4mm]
K= (-)^N \qquad K a = - a K .
\end{array}
\end{equation} 
 In this equation $K$ is the exchange operator ( see Sections 3 and 4 ) which here acts  simply as a parity operator 
 that separates the set of excited 
 states $|n\rangle \propto a^{\dagger n}|0\rangle $ into even and odd subspaces.
For $\nu > - \frac{1}{2}$,  the algebra (3) possesses 
unitary infinite - dimensional representations.
This algebra is known as the Calogero-Vasiliev algebra [6] ( also termed the deformed Heisenberg algebra with
reflection [12] ) and 
 provides an algebraic formulation of the two - 
 particle Calogero model  [1] 
described by   the Hamiltonian ( $x$ and $p$ are the relative coordinate and momentum, respectively )
\begin{equation}
2 H= p^2 + x^2 +\frac{\nu (\nu -1)}{x^2} K 
\end{equation}  
 which reduces to 
\begin{equation} 
 2 H = \{a,a^{\dagger}\} 
\end{equation} 
 after the identification
$$
\sqrt {2} a = x + i p - \frac{\nu }{x} K\\[4mm]
$$
$$
\sqrt {2} a^{\dagger} = x - i p + \frac{\nu }{x} K .
$$
 
{\it Remark \, 1.} Generalizations of the algebra (3) have been investigated in Refs. [13,14] and its connection to nonlinear
parabosonic (parafermionic) supersymmetry have been described in [15].
 
As we have already described in [7,9], the  analysis of the general ( deformed ) oscillator algebras  could be 
carried out in three
 completely equivalent ways.\\
 One can express $aa^{\dagger}$ as a normally ordered expansion: 
 \begin{equation} 
 aa^{\dagger}= 1+ \sum_{k\geq 1}\alpha_k \; a^{\dagger k}\;a^{k}.
 \end{equation} 

 Alternatively, one can  start the analysis by using the action of the annihilation operator on the states 
in Fock space:
\begin{equation} 
a a^{\dagger m}|0\rangle = \phi (N+1)a^{\dagger m-1}|0\rangle .
\end{equation}

 The third way is to examine   the vacuum matrix elements ( Gram matrix )
\begin{equation}
A_{m,n}=\langle 0|a^m a^{\dagger n}|0\rangle =[\phi(n)]!\;\delta_{mn}.
\end{equation}
These approaches are  rather simple for the single-mode oscillators but becomes very  powerfull 
for the analysis of multimode (deformed) oscillator algebras (see Sections 3,4.).

Now, we  show how the normally ordered expansion (6) works for  the Calogero-Vasiliev algebra (3).\\
 First,   
we  calculate the function $\phi(N)$  and it reads
\begin{equation}
\phi (N)=N +\nu \,( 1 + (-)^{1+N} ).
\end{equation}
Knowing $\phi(N)$, we  recursively calculate  the coefficients $\alpha_k$ ( Eq.(6)) for the algebra (3):
 $$
 \alpha_k=\frac{\phi (k+1)-1-\sum_{m=1}^{k-1}\alpha_m \;\phi(k)\cdots \phi(k+1-m)}
{[\phi(k)]!},\qquad \forall \phi(k) \neq 0. 
 $$
Similarly, we can expand the operators $K$ and $N$ 
 in an infinite series in operators 
$a$ and $a^{\dagger}$, i.e.
\begin{equation}
K=1 + \sum_{k \geq 1}\, \beta_k\;  a^{\dagger k} \, a^k .
\end{equation}
Using relations (3) and $Ka^{\dagger n}|0\rangle = (-)^na^{\dagger n}|0\rangle $,
 one can recursively calculate the coefficients $\beta_k$ as
$$
\beta_k=\frac{[(-)^k -1]}{\phi(k)!} - \sum_{m=1}^{k-1}\beta_m \;\frac{1}{\phi(m-1)!} .
$$

The expansion of the  number operator $N$ reads
\begin{equation}
N= a^{\dagger}a+\sum_{n\geq 2}\gamma_n \, (a^{\dagger})^n(a)^n,
\end{equation}
where
$$\gamma_n=\frac{n-\sum_{k=1}^{n-1} \gamma_k \; \phi(n)\cdots \phi(n+1-k)}
{[\phi(n)]!}.
$$
Note that $\gamma_n=0$ for $\phi(n)=0$, $\phi(n-1)\neq 0$.\\
Notice that,  knowing $\alpha_k$, $\beta_k$  and $\gamma_n$, we can
obtain $\phi(n)$ from the same recurrent relation.

As we elaborated in [9], there exists a simple mapping of the general deformed 
algebra (1) 
to the ordinary Bose algebra $[ b, b^{\dagger}]=1$. The mapping is of the form 
\begin{equation}
a = b \sqrt {\frac{a^{\dagger} a}{N}}\equiv b\sqrt {\frac{\phi(N)}{N}} .
\end{equation} 
 The inverse mapping exists if
$ \phi(N) \neq 0$, i.e.  $ \phi(N) > 0$, $\forall N$. \\  
Using $\phi (N)$, Eq.(9), associated with the  algebra (3), we obtain
$$
a=\left \{ \begin{array}{ll}
b  & \mbox { $n = even$}\\
b\sqrt {\frac{N+2\nu}{N}}& \mbox { $n = odd$} .\\
\end{array}
\right.
$$
It is also possible to map the operators $a$ and $a^{\dagger}$ to the Bose operators $b$ and $b^{\dagger}$, using 
 the expansion of the form 
\begin{equation}
a= (\; \sum_{k \geq 0} \, c_k \; b^{\dagger k}\,   b^k  \;)\cdot b .
\end{equation} 
Comparing Eq.(13) with Eq.(3), one can recursively calculate the coefficients $c_k$ as
$$
c_k=\frac{1}{k!} [ \sqrt {\frac{\phi(k+1)}{(k+1)}} + \sum_{m=1}^k (-)^k 
\left ( \begin{array}{c}
k\\m
\end{array} \right )
\sqrt {\frac{\phi(k+1-m)}{(k+1-m)}} ] .
$$

For  further purpose ( see Section 4), it is convenient to define a new 
operator $\tilde{a} $ in a sense dual to $(a , a^{\dagger}$),  such that 
\begin{equation}
[ \tilde{a}, a^{\dagger}]=1, \qquad  \tilde{a}|0\rangle =0.
\end{equation} 
The connection  is
\begin{equation}
\tilde{a}= a \,\frac{N}{a^{\dagger} a}\equiv a\,\frac{N}{\phi (N)},\qquad  \phi(N) > 0, \qquad \forall N
\end{equation} 
where $N$ can be realized as in (11), or  as
$$
N=\frac{1}{2} \{a^{\dagger},a\} - (\nu +\frac{1}{2}) .
$$
Using $\phi (N)$, Eq.(9), we find
\begin{equation}
\tilde{a}=\left \{ \begin{array}{ll}
a  & \mbox { $n = even$}\\
a\frac{N}{N+2\nu}& \mbox { $n = odd.$}\\
\end{array}
\right.
\end{equation}
The construction of the  mapping (15) is  
similar to that given in [16].\\
We also find
$$
N=a^{\dagger}\tilde{a}, \qquad N+1=\tilde{a}a^{\dagger},
$$
and 
$$
\tilde{a}a^{\dagger n}|0\rangle =n a^{\dagger (n-1)}|0\rangle 
\qquad \langle0|\tilde{a}^m a^{\dagger n}|0\rangle=n!\,\delta_{mn}.
$$

In the next section we turn to  general multi-mode oscillator algebras with permutational invariance.



 \section { Intermezzo: Multi-mode oscillator algebras with permutation
 invariance }
As we emphasized in [7,8], the analysis of the general multimode ( deformed )
oscillator algebras is more complicated and the three approaches, Eqs.(6-8), 
become more involved. Here we  concentrate on permutation invariant
multi-mode oscillator algebras. Invariance on the permutation group simplifies the 
analysis but
 one still has to distinguish  between two cases. The first case are 
permutational invariant algebras with hermitean number operators. These algebras
were analysed in [7,8]. In order to be a self-contained, in  Subsection 3.1
 we repeat the main points of this analysis.\\
The second case are permutation invariant algebras with non-hermitean 
number operators, not discussed previously.
 The analysis of these algebras is presented in  Subsections 3.2 and 3.3. 

\subsection{ Permutation invariant algebras with hermitean number operators}

Let us consider a system of multi-mode oscillators described by $M$ pairs of 
creation and annihilation operators $a_{i}^{\dagger}$, $a_{i}$ ($i = 1,2,...,M$) 
hermitian conjugated to each other.
We consider operator algebras which possess  the well-defined transition 
number operators $N_{ij}$, the partial number operators $N_i\equiv N{_{ii}}$ and  
the total number operator  $N=\sum N_i$. We also 
demand that the algebras be permutation invariant.
In this subsection we suppose that all number operators are ${\it hermitean }$,
 i.e. $N_i^{\dagger}=N_i$, 
$N_i^{\dagger}=N_i$ and $N_{ij}^{\dagger}=N_{ji}$.\\
The relations involving the number operators
and the operators $a_{i}^{\dagger}$, $a_{i}$ ($i = 1,2,...,M$) are 
$$
[N_{ij},a_k^{\dagger}]= \delta_{jk} a_i^{\dagger}, \qquad
[N_{ij},a_k]= -\delta_{ik} a_j, \qquad
[N_{ij},N_{kl}]=\delta_{jk}N_{il}-\delta_{il}N_{kj}.
$$
$$
[N_i,a_j]= -\delta_{ij} a_i, \qquad [N_i,a_j^{\dagger}]= 
\delta_{ij} a_i^{\dagger},\qquad [N_i,N_j]=0
$$
\begin{equation}
[N,a_k^{\dagger}]= a_k^{\dagger},\qquad [N,a_k]= -a_k.
\end{equation}

In the associated Fock-like representation, let $|0\rangle$ 
denote the vacuum vector. Then, the scalar product is uniquely defined by $\langle 0|0\rangle = 1$, 
and the vacuum 
conditions are $a_{i}|0\rangle = 0$, $a_i a_i^{\dagger}|0\rangle \neq 0$. A general $n$-particle 
state is a linear combination of monomial state vectors 
$(a^{\dagger}_{i_{1}}\cdots a^{\dagger}_{i_{n}}|0\rangle)$, 
$i_{1},...,i_{n} = 1,2,...,M$. The partial number operators $N_i$ are diagonal on the monomial states 
($a^{\dagger}_{i_1} \cdots a^{\dagger}_{i_n}| 0 \rangle $), with eigenvalues $n_i$, counting the number of 
operators $ a^{\dagger}_i$ in the corresponding monomial state:
$$
N_i (a^{\dagger}_{i_1} \underbrace{a^{\dagger}_i \cdots a^{\dagger}_i}_{n_i} a^{\dagger}_{i_n}| 0 \rangle ) = n_i 
(a^{\dagger}_{i_1} \underbrace{a^{\dagger}_i \cdots a^{\dagger}_i}_{n_i} a^{\dagger}_{i_n}| 0 \rangle ).
$$
In  Fock space considered, monomial states with different total number operator $N$ are 
orthogonal, as well as the states with the same $N$ but different partial 
number operators $N_i$. 
Two states are not orthogonal if they have the same partial number operators 
 $N_1, N_2,...N_M$. \\
The general multi-mode oscillator algebra with hermitean number operators and permutation invariance 
can be described in three  ways [7]: (i) as a normally 
ordered expansion, (ii) as a set of  relations in the Fock space and,  (iii) using Gram matrix of
scalar products in  Fock space. These three approaches are independent but completely equivalent. Moreover, as we
demonstrate in the rest of the paper, they can
also be applied to the permutation invariant algebras with non-hermitean number operators.
 Which of the three approaches  will be used,
depends on the nature of the problem. In the  concrete example of the Calogero model ( Section 4 ) we are mainly 
interested in positivity of the norms in  Fock space and in finding the critical points. Therefore, we shall analyse 
the problem using the Gram matrix.\\
Now, we  shortly describe each of the three approaches.

In the first approach, the operator algebras are defined by a set of relations:
$$
a_{i} a_{j}^{\dagger} \equiv \Gamma_{ij}(a^{\dagger},a) =  
$$
\begin{equation}
=\delta_{ij} + C_{1,1} \;a^{\dagger}_{j} a_{i} 
+ \sum_{n=1}^{\infty}\,\sum_{\pi,\sigma \in S_{n+1}}\;\;C_{\pi,\sigma}\,
\sum^{M}_{k_{1},...,k_{n} =1}\,[\pi(j,k_{1},...,k_{n})]^{\dagger}\,[\sigma(
i,k_{1},...,k_{n})]\,,
\end{equation}
where the operators $a_{i}$ are normalized in such a way that the 
coefficient of the $\delta_{ij}$ term is equal to $1$. Several comments on the structure of  
 the above expression are in order . \\
Permutation invariance guarantees that the coefficients in the expansion do not depend on concrete indices
 in normally ordered monomials, 
but only on certain linearly independent types of permutation invariant terms, schematically displayed above.
The existence of the number operators $N_i$ implies that the annihilation 
and creation operators appearing in a monomial in the normally ordered expansion 
(18) have to appear in pairs, i.e. monomials are diagonal in the variables
$k_1,...,k_n$ (up to permutations) [8]. 
The symbol $[\sigma(i,k_{1},...,k_{n})]$ 
denotes $a_{\sigma(i)}a_{\sigma(k_{1})}\cdots a_{\sigma(k_{n})} 
\equiv \sigma(a_{i}a_{k_{1}}\cdot\cdot\cdot a_{k_{n}})$. Also, 
$C_{\pi,\sigma} = C^{\ast}_{\sigma,\pi}$, owing to the hermiticity of the 
operator product $a_{i}a^{\dagger}_{i}$. 
We consider only those relations in (18) that may allow the norm zero vectors, but 
do not allow the state vectors of negative norm in  Fock space. 
 The norm zero vectors imply 
relations between  creation (annihilation) operators. Since  these relations are 
consequences of Eq.(18), they need not be postulated independently. Also , there is no need to
 postulate relations [ $a_ia_i^{\dagger}$ ] separately, since they can differ from the relations 
 [ $a_ia_j^{\dagger}|_{i=j}$ ] only in the unique function $f(N,N_i)$.
Finally, notice that,  although there are infinitely 
many terms in the expansion, only finite terms are actually involved when 
acting on the finite monomial state in  Fock space.

In the second approach, in addition to the vacuum relations $a_{i}|0\rangle = 0$,\\ 
$a_i a_j^{\dagger} | 0 \rangle = \delta_{ij}|0 \rangle $, 
one can define the action of $a_i$, $i=1,2,...M$, on monomial states 
$( a^{\dagger}_{i_1} \cdots a^{\dagger}_{i_n}| 0 \rangle )$ through  the relations 
\begin{equation}
a_{i}a^{\dagger}_{i_{1}}\cdots a^{\dagger}_{i_{n}}|0\rangle = \sum^{n}_{k=1}\,
\delta_{i i_{k}}\,\sum_{\sigma \in S_{n-1}}\,
\phi^{k}_{\sigma}[\sigma(i_1,..,\hat{i}_k,..,i_{n})^{\dagger}]\,|0\rangle\,,
\end{equation}
where $\hat{i}_{k}$ denotes the omission of the creation operator ( with index $i_k$ ) in all possible ways.
( If the monomial state does not contain $ a^{\dagger}_{i}$ at all, the RHS of (19) is equal to zero .)
The sum is running over all linearly independent monomials and 
$\phi^k_{\sigma}$ are (complex) coefficients.
 The identity $\phi^1_{id} = 1$ is implied by 
normalization in Eq. (18). The coefficients $\phi^k_{\sigma}$ can be 
uniquely determined from $C_{\pi,\sigma}$ and vice versa.
If we write the type of monomial state $( a^{\dagger}_{i_1} \cdots a^{\dagger}_{i_n}| 0 \rangle )$ as
$1^{\nu_1}2^{\nu_{2}}...M^{\nu_{M}}$, where $\nu_{1},\nu_{2},...,\nu_{M}$ are multiplicities 
satisfying $\nu_{i}\geq 0$ and $\sum^{M}_{i=1} \nu_{i} = n$, we see that 
permutation invariance drastically reduces the number of independent terms in Eq.(19), 
i.e. from $M^{n+1}$ \\
( for the general algebra ) to at most 
( $\sum_{k=1}^n\; k \frac{n!}{(\nu_{1}!\cdots \nu_{k}!)}$ ) independent terms ( for the permutation invariant algebra ).

{\it Example 1.} States for $n=2$: hermitean $( N_{ij},N_i )$\\
$$
a_1(a_1^{\dagger})^2| 0 \rangle
$$
$$
a_1a_1^{\dagger}a_2^{\dagger}| 0 \rangle, \qquad a_2a_1^{\dagger}a_2^{\dagger}| 0 \rangle.
$$
In the third approach, one defines the Gram matrix of scalar products 
\begin{equation}
A_{i_n\cdots i_1;j_1 \cdots j_n}=\langle 0 |a_{i_n} \cdots
a_{i_1}a^{\dagger}_{j_1}\cdots a^{\dagger}_{j_n}| 0 \rangle . 
\end{equation}
For the permutation invariant algebra, it is
permutation invariant, meaning that the matrix element
$ \langle 0|a_{i_{\pi (n)}} \cdots a_{i_{\pi (1)}}
a^{\dagger}_{j_{\pi (1)}} \cdots a^{\dagger}_{j_{\pi (n)}}|0\rangle $ does not 
depend on the permutation $\pi \in S_n $. The rank of the Gram matrix gives the number of linearly independent
states in  Fock space, which is positive definite if $A\geq 0$, i.e. if all eigenvalues are non-negative.
 The matrix and its rank depend only on the collection of multiplicities
 $\{\nu_1,...,\nu_M\}$, which, written in descending order, give rise to a partition of $n$ [8].
The generic matrix ( all indices different! ) is of the type  $n!\times n!$. It can be decomposed into terms
of the right regular representation of the permutation group $S_n$ [8]. All other, non-generic
matrices,   are easily obtained from the generic matrix. Their order is  
$\frac{n!}{\nu_1 !\cdots \nu_k !}\times\frac{n!}{\nu_1 !\cdots \nu_k !}$, where $\sum \nu_k =n$.\\
We mention that the typical permutation invariant algebras with hermitean number operators are 
parastatistics/interpolation between parastatistics [17] and infinite quon statistics [18].
Finally, we note that there is a simple way to unify a very large class of such  permutation invariant
algebras as a triple operator algebras [19]:
\begin{equation}
[\,[a_i,a^{\dagger}_j]_q, a^{\dagger}_k\, ]=x \delta_{ij}a_k^{\dagger} + y\delta_{ik}a_j^{\dagger}
+ z\delta_{jk}a_i^{\dagger}, \qquad (x,y,z) \in \bf {R}
\end{equation}
which can be rewriten as [20]
$$
 a_ia_j^{\dagger} = q a_j^{\dagger}a_i + ( 1+x N )\delta_{ij} + y N_{ij} + z N_{ji}.
$$
\subsection{ Permutation invariant algebras with non-hermitean \\number operators}

There exist a large class of operator algebras which possess permutation invariance  but for these 
$N_i\neq N_i^{\dagger}$ and $N_{ij}^{\dagger}\neq N_{ji}$ ( it still holds  $N=N^{\dagger}$ ). The most 
important example is the many-body Calogero model[1,6]. 
Generally, in these algebras one can have $a_i|0\rangle =0$, but 
$a_ia_j^{\dagger}|0\rangle \neq 0$ for $i \neq j$. As a
rule, a non-orthogonal monomial basis  appears, i.e.   two monomial states with the same total 
number operator $N$  but different partial number operators $N_i$ are not orthogonal.\\
 The whole algebra can be obtained from one relation, e.g. $a_1a_2^{\dagger}$, and by successive 
application of 
permutation $\pi \in S_M$, $a_{\pi(1)}a^{\dagger}_{\pi(2)} = \pi (a_1a_2^{\dagger})$. The general structure
of the normally ordered expansion is
\begin{equation}
a_ia_j^{\dagger}= c_0 + [a_j^{\dagger}a_i] + [a_j^{\dagger}B_{0,1} + B_{0,1}^{\dagger}a_i] +
  [B_{0,1}^{\dagger}B_{0,1}] + [B_{1,1}] + \cdots  ,
\end{equation}
where
$$
  B_{m,n}=\sum_{k=1}^M (a_k^{\dagger})^m a_k^n \qquad B_{m,n}^{\dagger}=B_{n,m}.
$$
The above expansion displays the $S_M$ symmetric structure as it contains the $S_M$-symmetric 
operators only. Notice that, although the RHS of Eq.(22) contains an equal number of $a^{\dagger}$'s and $a$'s, 
they are not matched in pairs any longer (cf. Eq.(18)). The general structure of the terms in the expansion of
 $a_ia_j^{\dagger}$ in (22) ( as well as the structure of $N_{ij}$, $N_i$ and $K_{ij}$, see later ) is
 $$
 a_m^{\dagger k}\; {\cal {O}} \; a_n^l, \qquad m,n = i,j  
 $$
 where ( $a_m$, $a_n$ ) are $a_i$ or $a_j$ and $\cal O$ is any normally ordered $S_M$-invariant polynomial 
 in the operators $( a^{\dagger}, a )$. The indices $(i,j)$ appear explicitly in the expansion and all other 
 indices are contained implicitly in the  $S_M$-invariant operator $\cal O$.
  Using this fact, one can uniquely define   $[ a_ia_j^{\dagger}|_{i=j} ]$ and it 
should coincide with [ $a_ia_i^{\dagger}$ ] up to the unique hermitean function of the operators 
$(N,N_i,N_i^{\dagger},N_{ik},..)$. which are no longer diagonal on monomial states  ( and 
can change the eigenvalues of $N_k$ for fixed total N).

Going to the Fock-space description (19), we notice that, owing to the non-orthogonality
 of monomial states, there appear much more independent terms than in the orthogonal case (see Example 1).

{\it Example \,2}: States for $n=2$: non-hermitean $( N_{ij},N_i )$\\
$$
a_1(a_1^{\dagger})^2| 0 \rangle, \qquad a_2(a_1^{\dagger})^2| 0 \rangle
$$ 
$$
a_1a_1^{\dagger}a_2^{\dagger}| 0 \rangle,\qquad a_2a_1^{\dagger}a_2^{\dagger}| 0 \rangle,
\qquad a_3a_1^{\dagger}a_2^{\dagger}| 0 \rangle. 
$$
Generally, there are at most 
 $$
 \sum_{k=1}^{n} (k+1)\; \frac{n!}{\nu_1 !...\nu_k!}, \qquad \sum_{i=1}^k\nu_i=n
 $$
 independent terms.\\
The Gram matrices $A_{i_n\cdots i_1;j_1 \cdots j_n}$ are hermitean,  of the type $M^n$ and we require 
all eigenvalues to be non-negative.
 The matrix elements of the particular Gram matrix are related by permutation symmetry. In the next section we 
  study 
 the structure of these matrices in more detail. 

\subsection{Operators $N_{ij}$, $N_i$ and $K_{ij}$}

The transition number operators $N_{ij}$  can be expanded into an infinite
( normally ordered ) series in creation and annihilation operators. However, 
only finitely many terms are involved when $N_{ij}$ acts on a finite monomial state in 
Fock space.\\
We find that the general structure of $N_{ij}$ is (cf. (22))
\begin{equation}
N_{ij} = [a_i^{\dagger}a_j] +  [a_i^{\dagger}B_{0,1} + B_{0,1}^{\dagger}a_j] + [B_{0,1}^{\dagger}B_{0,1}]
 +[B_{1,1}] +\cdots
\end{equation}
The partial number operators $N_i$ are obtained from the above expression as  $N_i=N_{ii}$.
 Notice that $N_{ij}^{\dagger} \neq N_{ji}$.\\ 
The structure of the total number operator $N=\sum N_i$ is
\begin{equation}
N = [B_{0,1}^{\dagger}B_{0,1}] + [B_{1,1}] + [B_{0,2}^{\dagger}B_{0,2}] + [B_{0,2}^{\dagger}B_{0,1}^2] + 
[B_{0,1}^{\dagger 2}B_{0,2}] + [B_{0,1}^{\dagger 2} B_{0,1}^2 ]+ 
\end{equation}
$$
+[B_{2,1}B_{0,1}] + [B_{0,1}^{\dagger}B_{1,2}] + [B_{2,2}] + \cdots
$$
As we are interested only in  the $S_M$- symmetric structure,  the coefficients in the above 
expressions have been omitted. Notice that $N^{\dagger}=N$.

The exchange operators $K_{ij}$ , $i,j = 1,2,...M$, generate the symmetric group $S_M$. They are defined
as follows:
$$
K_{ij}=K_{ji}, \; (K_{ij})^2=1, \; K_{ij}^{\dagger}=K_{ij},
$$
\begin{equation}
K_{ij}K_{jl}=K_{jl}K_{il}=K_{il}K_{ij}, \qquad i\neq j, \; i\neq l,\;j\neq l.
\end{equation}

The representation of the symmetric group $S_M$ exists on every deformed algebra of $a_i$ and 
$a_i^{\dagger}$, ($i=1,....M$), in the sense
\begin{equation}
K_{ij}a_j=a_iK_{ij},\qquad K_{ij}a_k=a_kK_{ij},
\end{equation}
$$
K_{ij}a_j^{\dagger}=a_i^{\dagger}K_{ij},\qquad K_{ij}a_k^{\dagger}=a_k^{\dagger}K_{ij},\\
$$
for $k\neq i$ and $k\neq j$.\\
 The vacuum condition is $K_{ij}|0\rangle=\pm |0\rangle$,
and we choose $K_{ij}|0\rangle= + |0\rangle$.\\

${\it Remark \, 2}$.
Note that the following change of the definitions in (26):
$$
K_{ij}a_j= - a_iK_{ij},\qquad K_{ij}a_k= + a_kK_{ij},\\
$$
(and similarly for $a_i^{\dagger}$ ), directly leads to contradiction since, from Eqs.(25,26), we 
obtain two apparently different results:
$$
K_{ij}K_{jk}a_k = a_iK_{ij}K_{jk}
$$
$$
K_{jk}K_{ki}a_k = - a_iK_{jk}K_{ki}.
$$

The important fact is that if the algebra of the oscillators
 is permutation invariant, then the $K_{ij}$ operators 
can be expressed similarly as $N$, $N_i$,  $N_{ij}$, namely as an  infinite series expansion in 
creation and annihilation operators.
 If the algebra  is not permutation invariant,
 such a representation of exchange operators  may not exist .
We point out the difference between the operators  $N_{ij}$ and $K_{ij}$. The exchange operators
$K_{ij}$ act "globally" ( and simultaneously ) on the right on any monomial in $a^{\dagger}$ and
$a$, interchanging indices $i$ and $j$ and keeping all other indices at rest.
Transition number operators $N_{ij}$ act "locally", turning only one $a_j^{\dagger}$ into 
$a_i^{\dagger}$, at the same place where $a_j^{\dagger}$ is sitting. The action of $N_{ij}$ 
can be repeated at most  $n_j$ times, where $n_j$ counts the number of $a_j^{\dagger}$'s in the monomial. 
If there is only one $a_j^{\dagger}$, then we have
$$
N_{ij} (\cdots a_j^{\dagger}\cdots ) |0\rangle =  (\cdots a_i^{\dagger}\cdots ) |0\rangle .
$$
Let $[a_i^{\dagger},a_j^{\dagger}]=0$ for $i\neq j$. We symbolically denote the eigenstate of $N_i$ as\\
$( \cdots a_i^{\dagger {n_i}} \cdots a_j^{\dagger {n_j}}\cdots )|0\rangle $. Then,
$$
N_i ( \cdots a_i^{\dagger {n_i}} \cdots a_j^{\dagger {n_j}}\cdots )|0\rangle =
n_i( \cdots a_i^{\dagger {n_i}} \cdots a_j^{\dagger {n_j}}\cdots )|0\rangle 
$$
$$
N_j ( \cdots a_i^{\dagger {n_i}} \cdots a_j^{\dagger {n_j}}\cdots )|0\rangle =
n_j ( \cdots a_i^{\dagger {n_i}} \cdots a_j^{\dagger {n_j}}\cdots )|0\rangle 
$$
$$
N_{ij}( \cdots a_i^{\dagger {n_i}} \cdots a_j^{\dagger {n_j}}\cdots )|0\rangle =n_j 
( \cdots a_i^{\dagger {n_i+1}} \cdots a_j^{\dagger {n_j-1}}\cdots )|0\rangle 
$$
$$
N_{ij}^{n_j}( \cdots a_i^{\dagger {n_i}} \cdots a_j^{\dagger {n_j}}\cdots )|0\rangle =n_j !
( \cdots a_i^{\dagger {n_i+n_j}} \cdots a_j^{\dagger 0}\cdots )|0\rangle 
$$
$$
N_{ji}^{n_i}N_{ij}^{n_j}( \cdots a_i^{\dagger {n_i}} \cdots a_j^{\dagger {n_j}}\cdots )|0\rangle =
(n_i+n_j) !
( \cdots a_i^{\dagger {n_j}} \cdots a_j^{\dagger {n_i}}\cdots )|0\rangle 
$$

 We also obtain
\begin{equation}
K_{ij}=\frac{1}{(n_i + n_j)!}( N_{ji} )^{n_i}( N_{ij} )^{n_j}=\frac{1}{(n_i + n_j)!}
 ( N_{ij} )^{n_j}( N_{ji} )^{n_i},
\end{equation}
or alternatively
$$
K_{ij}=\left \{ \begin{array}{ll}
( N_{ij}\frac{1}{N_j} )^{n_j - n_i} & \mbox {if $n_i < n_j$}\\
1 & \mbox {if $n_i = n_j$}\\
( N_{ji}\frac{1}{N_i} )^{n_i - n_j} & \mbox {if $n_i > n_j$}
\end{array}
\right.
$$
where $n_i$ and $n_j$  again denote, respectively, the number of $a_i^{\dagger}$ and $a_j^{\dagger}$ 
in the monomial $( \cdots a_i^{\dagger} \cdots a_j^{\dagger} )|0\rangle$. 
If $[a_i^{\dagger},a_j^{\dagger}]\neq 0$ for $i\neq j$,  there is generally no such a simple relation between
$K_{ij}$ and $N_{ij}$.\\
Below we give two examples of $K_{ij}$ operators for permutation invariant algebras with hermitean 
number operators. An example of $K_{ij}$ operators for permutation invariant algebra 
with non-hermitean number operators is given in the next section.

${\it {Example \, 3}}$. Heisenberg algebra of Bose oscillators $b_i$, $b_i^{\dagger}$, $i=1,..M$.\\
 Algebra:
 $$
 [b_i,b_j^{\dagger}]=\delta_{ij}, \qquad [b_i^{\dagger},b_j^{\dagger}]=[b_i,b_j]=0.
 $$
Number operators:
$$
N_{ij}=b_i^{\dagger}b_j, \qquad N_i=b_i^{\dagger}b_i, \qquad N=\sum_{i=1}^M b_i^{\dagger}b_i.
$$
Exchange operators:
$$
K_{ij}= \;: e^{-(b_i^{\dagger}- b_j^{\dagger})(b_i-b_j)}: \; =\sum_{k=0}^{\infty} \frac{(-)^k}{k!}
(b_i^{\dagger}- b_j^{\dagger})^k (b_i-b_j)^k.
$$

${\it {Example \, 4}}$. Clifford algebra of Fermi oscillators $f_i$, $f_i^{\dagger}$, $i=1,..M$.\\
 Algebra:
 $$
  \{f_i,f_j^{\dagger}\}=\delta_{ij}, \qquad \{f_i^{\dagger},f_j^{\dagger}\}=\{f_i,f_j\}=0.
 $$
Number operators:
$$
N_{ij}=f_i^{\dagger}f_j, \qquad N_i=f_i^{\dagger}f_i, \qquad N=\sum_{i=1}^M f_i^{\dagger}f_i.
$$
Exchange operators:
$$
K_{ij}=: e^{-(f_i^{\dagger}- f_j^{\dagger})(f_i-f_j)}: =
1 - (f_i^{\dagger}- f_j^{\dagger})(f_i-f_j).
$$




\section{Application: The M - body Calogero model }
The results of a general analysis of permutation invariant multi-mode oscillator algebras 
with non-hermitean number operators ( Subsections 3.2 and 3.3 ) will be now applied to the 
M-body Calogero model. We also generalize a several concepts (e.g. an infinite series expansion of 
$a_ia_j^{\dagger}$, $N$, $N_i$, $N_{ij}$,  $K_{ij}$ and dual algebra) 
introduced in Section 2. We find,   as particularly 
interesting, the structure and the eigensystem of the Gram matrices. The analysis of the Gram matrices 
enable us 
to locate the universal critical point of the M - body Calogero model at $\nu =-\frac{1}{M}$.

\subsection{The M - body Calogero model and the multi-mode \\oscillator algebras}

The M - body Calogero model, describing M identical bosons on the line, is defined by 
the following Hamiltonian [1]:
\begin{equation}
H=-\frac{1}{2}\sum_{i=1}^M\partial_i^2+\frac{1}{2}\sum_{i=1}^Mx_i^2+\frac{\nu (\nu -1)}{2}
\sum_{i\neq j}^M\frac{1}{(x_i-x_j)^2}.
\end{equation}
For simplicity, we have set $\hbar$, 
the  mass of particles and the frequency of harmonic 
oscillators equal to one. The dimensionless constant $\nu$ is 
the coupling constant (and/or the statistical parameter) 
and $M$ is the number of particles. The Hamiltonian (28) can be factorized by creation and annihilation
operators of the $S_M$-extended Heisenberg algebra [6].\\
Let us introduce the following analogs of creation and annihilation operators [6]:
$$
a_i^{\dagger}=\frac{1}{\sqrt{2}}(-D_i+x_i),\qquad
a_i=\frac{1}{\sqrt{2}}(D_i+x_i),
$$
where 
$$ D_i=\partial_i+\nu \sum_{j,j\neq i}^M\frac{1}{x_i-x_j}(1-K_{ij})$$
are Dunkl derivatives and $K_{ij}$ are exchange operators (Eqs.(25-26)) 
generating the symmetric group $S_M$.
One can easily check that the commutators of creation and 
annihilation operators  are 
\begin{equation}
[a_i,a_j]=[a_i^{\dagger},a_j^{\dagger}]=0,
\end{equation}
$$
[a_i,a_j^{\dagger}]= A_{ij}=\left(1+\nu \sum_{k=1}^MK_{ik}\right)\delta_{ij}-\nu K_{ij}.
$$
 The action of $K_{ij}$ on $a_i^{\dagger}$ and $a_i$ is given in Eq.(26).

{\it Remark 3.} 
The following definitions 
$$
 K_{ij}f_j= - f_iK_{ij},
$$
$$
 K_{ij}f_k= - f_kK_{ij},
$$
are also consistent and one can study the algebra defined by the anticommutator $\{ a_i,a_j^{\dagger} \}$:
$$
\{a_i,a_j^{\dagger}\}= A_{ij}=\left(1+\nu \sum_{k=1}^NK_{ik}\right)\delta_{ij}-\nu K_{ij}
\qquad \forall i,j .
$$
Note that $\{ a_i,a_j\}\neq 0$, if $\nu \neq 0$. Moreover, one can show that the general form of the
algebra, namely $[a_i,a_j^{\dagger}]_q= A_{ij}(\nu )$, $|q|\leq 1$, with the condition 
$K_{ij}a_j=  a_iK_{ij}$ ,  Eq.(26), 
is equivalent to the algebra $[a_i,a_j^{\dagger}]_q= A_{ij}(-\nu )$, 
with the conditions $K_{ij}a_j= - a_iK_{ij}$ and  $K_{ij}a_k= - a_kK_{ij}$.

After performing a similarity transformation on the Hamiltonian $H$,\\ 
$  (\prod_{i<j}^M|x_i-x_j|^{-\nu})\; H \;(\prod_{i<j}^M|x_i-x_j|^{\nu})$  , 
we obtain the reduced Hamiltonian $H'$ which, when restricted to the space of symmetric functions,
takes the folowing simple form:
\begin{equation}
H'=\frac{1}{2}\sum_{i=1}^M\{a_i,a_i^{\dagger}\}=\sum_{i=1}^Ma_i^{\dagger}a_i+E_0 ,
\end{equation}
$$
[H',a_i]= - a_i, \qquad [H',a_i^{\dagger}]=  a_i^{\dagger} \qquad [H',\{a_i,a_j^{\dagger}\}]=0.
$$
Notice that one can define the general Hamiltonian $\bar{H}$ as
$$
\bar{H} =\frac{1}{2}\sum_{i=1}^M\{a_i,a_i^{\dagger}\}= N +E_0 ,
$$
which is not restricted to the symmetric states only, i.e it acts on the states in the whole Fock space. 
In the  space of symmetric states, it coincides with $H'$, $H'=\bar{H} $.
The ground-state energy is $E_0=\frac{M}{2}(\, 1+\nu (M-1)\,)$.
The Fock-space representation is defined by $ a_i |0\rangle =0$, $\forall i$ and 
$K_{ij}|0\rangle = \epsilon |0\rangle $, $\forall (i,j)$. As $\epsilon^2=1$, we fix 
$\epsilon $ to $+1$.
The Fock space is spanned by monomials $(a_1^{\dagger n_1}\cdots a_M^{\dagger n_M}
|0\rangle )$ . In the following we analyse the full Fock space of states since (i) we want to apply ideas from
Section 3 to the $S_M$ - extended Heisenberg algebra (29) and (ii) we want to obtain the positivity of physical states 
as a consequence of positivity of states in the complete Fock space.
Physical states are  symmetric ( antisymmetric ) states  for 
the bosonic ( fermionic ) systems. The algebraic analysis of the physical space of symmetric states for 
 a two- and a three-body Calogero model (28) is given in [21]. A general approach to the algebra of observables
and dynamical symmetry algebra for the M-body Calogero model was proposed in [22].

We point out that the algebra (29) can be defined in a new way, without exchange operators $K_{ij}$.
The construction relies on the generalization of the triple operator algebras (21) [19,20]. 
 The only difference is that now the number operators $N_i$ and $N_{ij}$
are not $\it hermitean $.\\
Eliminating $\nu K_{ij}=[a_i,a_j^{\dagger}]$ for $i\neq j$, we find
$$
[a_i,B_{0,1}^{\dagger}]=1, \qquad \forall i\\
$$
$$
a_i[a_i,a_j^{\dagger}]=[a_i,a_j^{\dagger}]a_j, \qquad \forall (i,j) \;i\neq j\\
$$
$$
a_i[a_j,a_i^{\dagger}]=[a_j,a_i^{\dagger}]a_j, \qquad \forall (i,j), \; i\neq j\\
$$
\begin{equation}
a_k[a_i,a_j^{\dagger}]=[a_i,a_j^{\dagger}]a_k, \qquad \forall (i,j,k) \;i\neq j\neq k \neq i.
\end{equation}
(It is understood that the hermitean counterparts of these relations also hold.)
Notice that the single-mode algebra (3) is a true triple operator algebra (21) since it can be rewritten as
$[\{a,a^{\dagger}\},a^{\dagger}]=2a^{\dagger}$ and $[\{a,a^{\dagger}\},a]=-2a$.\\
The algebra (31) is still a permutation invariant algebra, but the 
 indices on the LHS and RHS are not the same ( cf. the second and third relations in Eq.(31) ).
 The last relation in (31) can be written as
 $$
 [a_k,[a_i,a_j^{\dagger}]]=0, \qquad \forall (i,j,k), \;i\neq j\neq k \neq i\\
 $$
 $$
 [a_i,a_j^{\dagger}]=[a_j,a_i^{\dagger}], \qquad \forall (i,j). \\
 $$
 For the triple operator algebras,the Fock-space representation is defined by the
 two generalized vacuum conditions ( one can notice the similarity with Green's parastatistics [17] ):
 $$
 a_i|0\rangle=0, \qquad \forall i
 $$
 \begin{equation}
 a_ia_j^{\dagger}|0\rangle= -\nu |0\rangle,\qquad \forall (i,j), \;i\neq j
 \end{equation}
   Under these conditions,
  the Fock representation is uniquely determined and  is equivalent to the first
   construction, Eq.(29). However, this algebra does not depend on $\nu$ and $K_{ij}$, i.e. in the
   formulation of
    the triple operator algebras, the interaction
    parameter $\nu $ enters only through the vacuum condition (32).\\  
    We also obtain the consistency conditions in the Fock representation, 
    namely
$$
([a_i,a_j^{\dagger}])^2 =\nu^2 \qquad ,   [a_k,([a_i,a_j^{\dagger}])^2]=0, 
$$
\begin{equation}
[a_i,a_j^{\dagger}][a_j,a_k^{\dagger}]= [a_j,a_k^{\dagger}][a_i,a_k^{\dagger}] = 
[a_i,a_k^{\dagger}][a_i,a_j^{\dagger}].
\end{equation}
There is a simple generalization of the triple operator algebra to include 
fermions  (i.e. anticommutators)  and quons (i.e. q-commutators).
\subsection{Gram matrices for the Calogero model}
In the rest of this section, we  discuss the Calogero model in the framework of  the three
approaches proposed in  Section 3. The special attention is devoted to the Gram matrix approach.\\
In the first approach we express $a_ia_j^{\dagger}$ as a normally ordered expansion:
$$
a_ia_j^{\dagger}=-\nu K_{ij} + a_j^{\dagger}a_i \qquad \forall (i,j), \;i\neq j\\
$$
\begin{equation}
a_ia_i^{\dagger}=1 + a_i^{\dagger}a_i + \nu \sum_{l,l\neq i}K_{il} .
\end{equation}
This is obviously a permutation invariant algebra. We can write it in the form of Eq.(22)
 if we know the expansion of $K_{ij}$ in terms of $a_i$ and $a_j^{\dagger} $ (and vice versa).
 Later we shall give such a construction.
 
 In the second approach we have to know the action of $a_i$ on the monomial states 
 $ ( a_1^{\dagger n_1}\cdots a_M^{\dagger n_M}|0\rangle )$
in the Fock space.\\
  For the one - particle states $( a_i^{\dagger}|0\rangle )$  , using $
 a_i|0\rangle =0$, $ \forall i $, we find
 $$
 a_ia_j^{\dagger}|0\rangle = 
 \left \{ \begin{array}{ll}
 - \nu |0\rangle & \mbox {$i\neq j$}\\
(1 + (M-1))|0\rangle & \mbox {$ \forall i=j$}\\
\end{array}
\right.
$$
For the two - particle states $( a_i^{\dagger}a_j^{\dagger}|0\rangle\ )$ we find:
$$
a_ia_{j_1}^{\dagger}a_{j_2}^{\dagger}|0\rangle=[A_{ij_1}a_{j_2}^{\dagger}+
a_{j_1}^{\dagger}A_{ij_2}]|0\rangle .\\
$$
There are only four types of relations, owing to $[a_{j_1}^{\dagger},a_{j_2}^{\dagger}]=0$:
$$
a_1(a_1^{\dagger})^2|0\rangle = (\nu (M-2) +2) a_1^{\dagger} |0\rangle + \nu B_{0,1}^{\dag}|0\rangle ,\\
$$
$$
a_1(a_2^{\dagger})^2|0\rangle =  -\nu (a_1^{\dagger}+a_2^{\dagger})|0\rangle ,\\
$$
$$
a_1a_1^{\dagger}a_2^{\dagger}|0\rangle = (\nu (M-2) +1) a_2^{\dagger} |0\rangle ,\\
$$
$$
a_1a_2^{\dagger}a_3^{\dagger}|0\rangle = -\nu (a_3^{\dagger}+a_2^{\dagger}) |0\rangle .\\
$$

For the three - particle  states  $( a_i^{\dagger}a_j^{\dagger}a_k^{\dagger}|0\rangle\ )$ we find
$$
a_ia_{j_1}^{\dagger}a_{j_2}^{\dagger}a_{j_3}^{\dagger}|0\rangle=[A_{ij_1}a_{j_2}^{\dagger}
a^{\dagger}_{j_3}+a^{\dagger}_{j_1}A_{ij_2}a^{\dagger}_{j_3}+
a^{\dagger}_{j_1}a^{\dagger}_{j_2}A_{ij_3}]|0\rangle.
$$
There are seven types of relations:
$$
a_1(a_1^{\dagger})^3|0\rangle = (\nu (M-3) +3) a_1^{\dagger 2} |0\rangle + 
\nu ( a_1^{\dagger}B_{0,1}^{\dag} + B_{0,2}^{\dag} )|0\rangle ,
$$
$$
a_1(a_2^{\dagger})^3|0\rangle = -\nu (a_1^{\dagger}a_2^{\dagger}+a_1^{\dagger 2}+ 
a_2^{\dagger 2} ) |0\rangle ,
$$
$$
a_1(a_1^{\dagger})^2a_2^{\dagger}|0\rangle = (\nu (M-2) +2) a_1^{\dagger }a_2^{\dagger }|0\rangle
-\nu a_2^{\dagger 2}|0\rangle + \nu a_2^{\dagger}B_{0,1}^{\dag}|0\rangle ,
$$
$$
a_1a_1^{\dagger}(a_2^{\dagger})^2|0\rangle = (\nu (M-2) +1)a_2^{\dagger 2}|0\rangle - \nu 
a_1^{\dagger}a_2^{\dagger}  |0\rangle ,
$$
$$
a_1a_1^{\dagger}a_2^{\dagger}a_3^{\dagger}|0\rangle = (\nu (M-3) +1)a_2^{\dagger }a_3^{\dagger}|0\rangle ,
$$
$$
a_1(a_2^{\dagger})^2a_3^{\dagger}|0\rangle =  -\nu (a_2^{\dagger}a_3^{\dagger}+a_2^{\dagger 2}+ 
a_1^{\dagger}a_3^{\dagger} ) |0\rangle ,
$$
$$
a_1a_2^{\dagger}a_3^{\dagger}a_4^{\dagger}|0\rangle =  -\nu (a_2^{\dagger}a_3^{\dagger}+
 a_2^{\dagger}a_4^{\dagger} + a_3^{\dagger}a_4^{\dagger})|0\rangle .
 $$

It is a simple task to generalize these equations to an arbitrary multiparticle state, i.e. 
$ (( a_1^{\dagger} )^{n_1}\cdots (a_M^{\dagger})^{n_M})|0\rangle ) $: 

$$
a_1 ( a_1^{\dagger} )^{n_1}( a_2^{\dagger} )^{n_2} \cdots (a_M^{\dagger})^{n_M})|0\rangle 
\equiv a_1|n_1; n_2;\cdots  n_M \rangle =
$$
$$
n_1|n_1-1;n_2;n_3 \rangle + \nu \;sgn (n_1 - n_2) \sum_{k=1}^{|n_1 - n_2|} 
|min(n_1,n_2)+k-1; max(n_1,n_2)-k; n_3 \rangle +
$$
$$
+ \nu \;sgn (n_1 - n_3) \sum_{k=1}^{|n_1 - n_3|}
|min(n_1,n_3)+k-1; n_2; max(n_1,n_3)-k \rangle +\cdots \\
$$
\begin{equation}
+ \nu \;sgn (n_1 - n_M) \sum_{k=1}^{|n_1 - n_M|}|min(n_1,n_M)+k-1; n_2; \cdots ;max(n_1,n_M)-k \rangle
\end{equation}
We  use these formulas in the constructing  Gram matrices for different M\\
( the third approach ).

Now we easily obtain the structure of the matrix elements of Gram matrices\\
$\langle 0|a_{i_n}\cdots a_{i_1}a_{j_1}^{\dagger}\cdots a_{j_n}^{\dagger}|0\rangle$. We
explicitly give several examples ( for $M=2,3$ ) in the Appendix and below we discuss
 eigenvalues and eigenvectors of Gram matrices corresponding to  one - and two - particle
 states for any $M$.

(1) One-particle states ( $a_i^{\dagger}|0\rangle $, $i=1,2,...M $ ):\\
The matrix of  one-particle states is of order M and has only two distinct entries:
 $-\nu$ and $1+\nu (M-1)$. Its eigenvalues and typical eigenvectors are

$$
\begin{tabular}{|c|c|c|c|} \hline
Eigenvalue & Degeneracy & Eigenvector &  Comments \\ \hline \hline
1 & 1 & $B_{0,1}^{\dagger}|0\rangle $&  \\ \hline
$1+ M \nu $ &$ M-1$ &$ ( a_1^{\dagger} - a_i^{\dagger} )|0\rangle$ &$ i\neq 1$\\ \hline \hline
\end{tabular}
$$
The positivity condition implies that all eigenvalues should be positive, meaning \\$1+ M \nu >0$ or
$\nu > - 1/M$.

(2) Two-particle states ( $a_i^{\dagger}a_j^{\dagger}|0\rangle $, $(i,j)=1,2,...M$ ):\\
The matrix of  two-particle states is of order $M^2$ and has four distinct entries
 of the form
$$
\langle 0|a^2_1(a_1^{\dagger})^2|0\rangle =a \\
$$
$$
\langle 0|a^2_1(a_2^{\dagger})^2|0\rangle = \langle 0|a^2_1a_2^{\dagger}a_3^{\dagger}|0\rangle =
\langle 0|a_1a_2a_3^{\dagger}a_4^{\dagger}|0\rangle= b \\
$$
$$
\langle 0|a^2_1a_1^{\dagger}a_2^{\dagger}|0\rangle =\langle 0|a_1a_2a_1^{\dagger}a_3^{\dagger}|0\rangle = c \\
$$
$$
\langle 0|a_1a_2a_1^{\dagger}a_2^{\dagger}|0\rangle = d \\
$$
where $a= [1+\nu (M-1)][2+\nu (M-1)] - \nu^2 (N-1)$,  $b=- \nu -\nu^2 (M-2)$, $c=2 \nu^2$ and 
$d= [1+\nu (M-1)][2+\nu (M-2)]$.\\
Its eigenvalues and typical eigenvectors are

$$
\begin{tabular}{|c|c|c|c|} \hline
Eigenvalue & Degeneracy & Eigenvector &  Comments \\ \hline \hline
0 & $M(M-1)/2 $& $[a_i^{\dagger}, a_j^{\dagger}] |0\rangle $& 
$ \forall (i,j), i\neq j$\\ \hline
$2( 1+ M \nu )$ & M & $( \{ a_i^{\dagger}, B_{0,1}^{\dagger}  \}-2 B_{0,2}^{\dagger} )|0\rangle $&
$ \forall i , M\geq 2$\\ \hline 
$( 1+ M \nu )( 2+ M \nu )$& $M-1$& $( \{ (a_i^{\dagger} - a_1^{\dagger}),B_{0,1}^{\dagger} \} - M(a_i^{\dagger 2} -
a_1^{\dagger 2}))|0\rangle $&$i\neq 1; M \geq 3 $\\ \hline 
$2( 1+ M \nu )(1+\nu (M-1))$&$M(M-3)/2$ & $\{  (a_i^{\dagger} - a_j^{\dagger} ),
( a_k^{\dagger} - a_l^{\dagger} )\}|0 \rangle$ & $(i\neq j \neq k \neq l)$ \\ \hline \hline
\end{tabular}
$$
Here, $B_{0,1}=\sum_i a_i$ and $B_{0,2}=\sum_i a_i^2$. Note that null-eigenstates are identically equal to zero
owing to the commutation relation $[a_i^{\dagger}, a_j^{\dagger}]\equiv 0 $, which is satisfied $\forall (i,j)$.\\
The positivity condition implies again that all non-zero eigenvalues are positive, which is satisfied if 
$1+ M \nu >0$ , i.e. $\nu > - 1/M$.\\
 One can  show that the same condition for positivity of eigenvalues  
 for three and more particle states also holds. There is a universal critical point, $\nu = - 1/M$ , at which 
 all matrix elements of an arbitrary multi-state Gram matrix are equal to $ (k!/M^k )$, where k denotes a k-particle
 state. This can be proved by induction  and here  we sketch the proof.\\
   The generic Gram matrix is of type ( $M^k \times M^k$ ). At the critical point $\nu = - 1/M$  we find
 
$$
A_{ij}\equiv \langle 0| a_ia_j^{\dagger} |0 \rangle = \frac{1}{M} ,
$$
$$
A_{i_2i_1;j_1j_2}\equiv \langle 0| a_{i_2}a_{i_1}a_{j_1}^{\dagger} a_{j_2}^{\dagger}|0 \rangle =
$$
$$
=A_{i_1j_1} \langle 0|a_{i_2}a_{j_2}^{\dagger}|0 \rangle + 
A_{i_1j_2} \langle 0|a_{i_2}a_{j_1}^{\dagger}|0 \rangle = \frac{2}{M^2} ,
$$
$$
\vdots 
$$ 
$$ 
 A_{i_k\cdots i_1;j_1 \cdots j_k} \equiv 
 \langle 0| a_{i_k}\cdots a_{i_1} a_{j_1}^{\dagger}\cdots a_{j_k}^{\dagger} |0 \rangle = 
$$
$$
 = A_{i_1j_1} \langle 0| a_{i_k}\cdots a_{i_2} a_{j_2}^{\dagger}\cdots a_{j_k}^{\dagger} |0 \rangle  +
 A_{i_1j_2} \langle 0| a_{i_k}\cdots a_{i_2} a_{j_1}^{\dagger}a_{j_3}^{\dagger}\cdots a_{j_k}^{\dagger} |0 \rangle 
 + \cdots +
$$
$$
 + A_{i_1j_k} \langle 0| a_{i_k}\cdots a_{i_2} a_{j_1}^{\dagger}\cdots a_{j_{k-1}}^{\dagger} |0 \rangle =
$$
$$
 =k \frac{1}{M} \langle 0| a_{i_k}\cdots a_{i_2} a_{j_2}^{\dagger}\cdots a_{j_k}^{\dagger} |0 \rangle = \frac
 {k!}{M^k}  .
$$

The rank of this matrix is one. There is only one state, namely   
$B_{0,1}^{\dagger k}|0 \rangle $, which corresponds to the center of mass and has positive norm. The  corresponding 
eigenvalue is $\frac{k!}{M^k} M^k =k!$. All other eigenstates are null-states with zero norm .   
 Diagonal matrix elements are larger than $\frac{k!}{M^k}$ if $\nu > - 1/M$ 
 and the positivity conditions for eigenvalues are satisfied.
\bigskip   
 
{\it Remark 4}.
The critical point is universal in the sense that all algebras of the form $[a_i,a_j^{\dagger}]_q=A_{ij}(\nu )$,
$|q|\leq 1$ have the same critical point $\nu = - 1/M$. The case $q=-1$, 
for which the algebra takes a fermionic form, is of special interest:
$$
\{f_i,f_j^{\dagger}\}=A_{ij}, \qquad \{F,F^{\dagger}\}=M ,
$$
$$
\{f_i,F^{\dagger}\}=1, \qquad F^2= F^{\dagger 2}=0.
$$
Here, $F=\sum f_i $.  
It follows that the one-particle  Gram matrix for $q=-1$ is the same as for $q=1$: 
$$
\left(\begin{array}{cccc}
1+\nu (M-1) & -\nu & \cdots & -\nu \\
-\nu & 1+\nu (M-1) & \cdots & -\nu \\
\cdots & \cdots & \cdots & \cdots  \\
-\nu & -\nu &  \cdots & 1+\nu (M-1)
\end{array}
\right)
$$
The matrices for  two- 
and many-particle cases depend on $q$ and will be treated in separate paper. It appears that $\nu = - 1/M$ could be interpreted 
  as a physically interesting point [23]. At this point, the Fock space reduces to the Fock space of 
  a single harmonic oscillator, corresponding to the centre-of-mass coordinate.


\subsection{Operators $N_{ij}$, $N_i$, $N$  and $K_{ij}$}

Now, we proceed to the construction of $N_{ij}$, $N_i$, $N$  and $K_{ij}$ operators. This construction
can be performed for any $M$ but, for  simplicity, we present the results for the first non-trivial
case $M=3$. All these constructions exist only if the positivity condition , $\nu > -1/M$, is 
satisfied. For $M=3$, $\nu > -1/3$. 
The construction starts with expanding the corresponding operator in a
series in $a_i$ and $a_i^{\dagger}$; for example, (indices are omitted for brevity)
\begin{equation}
K_{ij}= c_0 + \sum c_1\, a^{\dagger}\,a +\sum c_2\,a^{\dagger}\,a^{\dagger}\,a\,a +\cdots
\end{equation}
Using the definitions (25,26), we act with (36) on the vacuum (which gives $c_0=1$), then on the one-particle state, 
the two-particle state,  etc... In this way, we obtain linear recursive relations which are easily solved. 
The result for $K_{12}$ and $M=3$ is
\begin{equation}
K_{12}= 1 -\frac{1}{(1+3\nu )} b_{12}^{\dagger}b_{12} + \frac{1}{2(1+3\nu )^2}b_{12}^{\dagger 2}b_{12}^2 - 
\frac{\nu }{2(1+3\nu )^2(2+3\nu )}b_{12}^{\dagger}b_{123}^{\dagger}b_{12}b_{123}+\cdots,
\end{equation}
where $b_{12}=a_1 - a_2$ and $b_{123} = a_1 + a_2 -2a_3$. One gets $K_{13}$ and 
$K_{23}$ from $K_{12}$ using permutation invariance. 
Knowing $K_{ij}$, one can  find a normally ordered expansion $a_ia_j^{\dagger}$.\\
Similarly, one finds ($M=3$)
$$
N_1=\frac{1}{(1+3\nu )}a_1^{\dagger}a_1 + \frac{\nu}{(1+3\nu )}a_1^{\dagger}B_{0,1} - 
\frac{\nu}{4(1+3\nu )(2+3\nu )} 
a_1^{\dagger}b_{231}^{\dagger}b_{23}^2 - 
$$
\begin{equation}
-\frac{\nu (1+\nu )}{4(1+3\nu )^2(2+3\nu )}a_1^{\dagger}b_{231}^{\dagger}b_{231}^2 -
\frac{\nu }{2(1+3\nu )^2(2+3\nu )}a_1^{\dagger}b_{23}^{\dagger}b_{23}b_{231}+\cdots.
\end{equation}
Here, $b_{23}=a_2 - a_3$ and $b_{231} = a_2 + a_3 -2a_1$. $N_{12}$ is easy to obtain from the above formula. 
Note that $N_1^{\dagger}\neq N_1$. Similarly, $N_{12}^{\dagger} \neq N_{21}$.  However, the total number operator
$N$ is hermitean, $N^{\dagger}=N$ and for $M=3$:
$$
N= \frac{1}{(1+3\nu )}\sum_{i=1}^3a_i^{\dagger}a_i + \frac{\nu}{(1+3\nu )}
(\sum_{i=1}^3a_i^{\dagger})(\sum_{i=1}^3a_i) + 
$$
$$ +\frac{\nu}{(1+3\nu )^2(2+3\nu )}\sum_{i<j=1}^3
(a_i^{\dagger} - a_j^{\dagger})^2(a_i - a_j)^2 +
$$
$$
+ \frac{2\nu ^2}{(1+3\nu )^2(2+3\nu )}
[\sum_{i=1}^3a_i^{\dagger 2} - \sum_{i<j=1}^3a_i^{\dagger}a_j^{\dagger}][\sum_{i=1}^3a_i 
- \sum_{i<j=1}^3a_ia_j]\equiv
$$
$$
\equiv \frac{1}{(1+3\nu )}B_{1,1} + \frac{\nu}{(1+3\nu )}B_{0,1}^{\dagger}B_{0,1} + 
$$
$$
+\frac{\nu }{(1+3\nu )^2(2+3\nu )}\{ 2\nu [ \frac{3}{2} B_{0,2}^{\dagger} - \frac{1}{2} B_{0,1}^{\dagger 2} ]
[ \frac{3}{2} B_{0,2} - \frac{1}{2} B_{0,1}^2 ]+ 
$$
\begin{equation}
+ 3 B_{2,2} + B_{0,2}^{\dagger}B_{0,2} - 
2 (B_{2,1}B_{0,1} + h.c.) + 2 \sum_{i=1}^3 a_i^{\dagger}B_{1,1}a_i \}
\end{equation}
The result is consistent with the general expression $\bar{H}-E_0 = 
\frac{1}{2}\sum_i \{a_i,a_i^{\dagger}\} -E_0 = N$. In the limit $\nu \rightarrow 0 $, we
reproduce the standard result $N=\sum_i a_i^{\dagger}a_i$. Although the above expressions seem to be
divergent at the critical point, it appears that for $\nu =- 1/3$,   the degrees of
freedom , the Fock space and the related algebra are substantially reduced [23] and the above expansions are completely regular,
giving
$N=\frac{1}{3}B_{0,1}^{\dagger}B_{0,1}$ at $\nu =- 1/3$.\\

\subsection{Dual operators $\tilde {a_i}$ and dual algebra $\tilde {\cal A}$ }
Owing to the fact that $[a_i,a_j]=0$ and $[a_i^{\dagger},a_j^{\dagger}]=0$, $\forall (i,j)$, 
we can define the operators $\tilde {a_i}$, $i=1,2,...M$, $\nu >-\frac{1}{M}$, such that
\begin{equation}
\tilde{a_i}(a_{i_1}^{\dagger }\cdots a_{i_m}^{\dagger }|0\rangle ) = \sum_{\alpha =1}^m\,
\delta_{ii_{\alpha }}\, a_{i_1}^{\dagger }\cdots \hat {a_{i_{\alpha}}^{\dagger }}\cdots 
a_{i_m}^{\dagger }|0\rangle ,
\end{equation}
$$
\tilde{a_i}|0\rangle =0 ,
$$
where the hat denotes omission of the corresponding operator. \\
The sum on the RHS contains $m_i$ terms. We find that:
$$
\tilde{a_i}\,(a_{i_1}^{\dagger m_1}\cdots a_i^{\dagger m_i}\cdots a_{i_M}^{\dagger m_M}|0\rangle )
= m_i (a_{i_1}^{\dagger m_1}\cdots a_i^{\dagger m_i-1}\cdots a_{i_M}^{\dagger m_M}|0\rangle )
$$
and
\begin{equation}
[\tilde{a_i},a_j^{\dagger}]=\delta_{ij}, \qquad  [\tilde{a_i},\tilde{a_j}]=0, \qquad \forall (i,j).
\end{equation}
These relations are satisfied on all monomial states in  Fock space. If we define a dual Fock space,
spanned by monomials $( \langle 0| \tilde{a}_{i_{1}}\cdots \tilde{a}_{i_{M}} )$, we obtain the following relation, as a
consequence of Eq.(41):
$$
\langle 0| \tilde{a}^{\tilde{m}_1}_{i_{1}}\cdots \tilde{a}^{\tilde{m}_M}_{i_{M}}
a_{i_1}^{\dagger m_1}\cdots  a_{i_M}^{\dagger m_M}|0\rangle =\prod_{k=1}^M n_k! \delta_{m_k\tilde{m}_k}
$$
We call the operators $\tilde{a_i}$ 
the bosonic duals of operators $a_i^{\dagger}$.\\
 The transition number operators $N_{ij}$,  the 
partial number operators $N_i$ and the total number operator $N$ can now be  expressed as:
$$
N_{ij}=a_i^{\dagger}\tilde{a_j}, \qquad \forall (i,j),
$$
$$
N_i=a_i^{\dagger}\tilde{a_i}, \qquad \forall i,
$$
$$
N=\sum_{i=1}^M a_i^{\dagger}\tilde{a_i} \qquad \forall i.
$$
From the expression for $N_{ij}$ we obtain $\tilde{a_j}$ and vice versa. Symbolically,
$$
\tilde{a_j}= a_j + \sum_{k\geq 1} a^{\dagger k}a^{k+1}.
$$
For example,  for  $M=3$ we find 
$$
\tilde{a_i}=\frac{1}{(1+3\nu )}a_i + \frac{\nu}{(1+3\nu )}B_{0,1} - \frac{\nu}{4(1+3\nu )(2+3\nu )} 
b_{kji}^{\dagger}b_{kj}^2 -
$$
$$-\frac{\nu (1+\nu )}{4(1+3\nu )^2(2+3\nu )}b_{kji}^{\dagger}b_{kji}^2 -
\frac{\nu }{2(1+3\nu )^2(2+3\nu )}b_{kj}^{\dagger}b_{kj}b_{kji} +\cdots,
$$
where $b_{kji}=(a_k - a_i)+ (a_j - a_i)$ and $b_{kj}=(a_k - a_j)$.
Hence, we obtain  new families of commuting operators $\tilde{a_i}(\nu )$, $i=1,2...M$ and 
$\nu >-1/M$. They satisfy a new commutation relation:
$$
[\tilde{a_i}(\nu ),\tilde{a_j}^{\dagger}(\nu )] = \tilde {A_{ij}}(\nu )
$$
and we call it dual algebra $\tilde {\cal A}(\nu )$ to the algebra of Eq.(29). Of course,
$[\tilde{a_i}(\mu ),\tilde{a_j}(\nu )]\neq 0$ for $\nu \neq \mu $ ( as $[a_i(\mu ),a_j(\nu )]\neq 0$
 for $\nu \neq \mu $ ).\\
  The definition and structure of the algebra dual to a general algebra of $a_i$ and $a_i^{\dagger}$ operators
is an interesting problem. Its physical meaning is connected with the construction of new models which
are dual to the initial one.

\subsection{Mapping to free Bose oscillators}
It was found that the M-body Calogero model in the harmonic potential (28) could be mapped 
to M ordinary free Bose
oscillators [24]. The mapping was performed in the coordinate space ( not in the Fock space ) and no
restriction on $\nu $ was found or discussed. Since the whole Fock space 
 (spanned by the monomials $a_{i_1}^{\dagger m_1}\cdots  a_M^{\dagger m_M}|0\rangle $) 
 for the M-body Calogero model with $\nu >-1/M$ , is isomorphic to the Fock space of M free Bose
  oscillators with $\nu=0 $, we conclude that there must exist a regular mapping from 
  $(a_i,a_i^{\dagger})$ to $(b_i,b_i^{\dagger})$ and vice versa. 
  To ensure the existence of the mapping
  $a=\Psi (b,b^{\dagger})$, the  following relations have to be satisfied.
  $$
  [a_i^{\dagger},a_j^{\dagger}]=[a_i,a_j]=0, 
$$
\begin{equation}
  [N_i,a_j^{\dagger}]=\delta_{ij}\, a_i^{\dagger},\qquad \forall (i,j).
\end{equation}
 The sufficient condition for the existence of the 
inverse real mapping $b=\Psi^{-1}(a,a^{\dagger})$ is $\nu >-1/M$.

Our results on mappings can be generalized in the following way. If two algebras of operators, e.g. 
$(a_i,a_i^{\dagger})$ and $(b_i,b_i^{\dagger})$, have completely isomorphic Fock spaces ( i.e. the same
structure of all Gram matrices ), then there exists a regular, real mapping from $a_i$ to $b_i$ and 
vice versa. If one Fock space is isomorphic with a subspace of the second Fock space, then there exists
the mapping $a=\Psi (b,b^{\dagger})$, but there is no inverse mapping. The construction of the
mapping for Calogero operators $a_i, i=1,..M$, (Eq.(29)), will be considered in a separate publication.



\section{Conclusion}
In conclusion, we want to point out the main results of the paper. In section 2 we have applied the general
results of Ref.[9] to the Calogero-Vasiliev single-mode oscillator algebra (3), underlying the two-body Calogero 
model. We have expressed the number operator $N$ (11) and the exchange operator $K$ (10) as an infinite series in creation
and annihilation operators and have  recursively calculated the coefficients of expansion. We have found 
a mapping (13)
from Calogero-Vasiliev oscillators $(a,a^{\dagger})$ to the Bose oscillators $(b,b^{\dagger})$. The mapping
has the form of an infinite series in $(b,b^{\dagger})$. Finally, we have defined a new operators 
$(\tilde{a}, \tilde{a}^{\dagger})$ which are dual to the operators $(a,a^{\dagger})$ in the sense that 
$[\tilde{a},a^{\dagger}]=1$. We have found a connection between the operators $a$, $\tilde{a}$ and $b$, 
Eqs.(12,15,16).\\
Section 3 is devoted to the generalization of the single-mode oscillator algebras to the multi-mode case. 
We have discussed two distinct classes of multi-mode oscillator algebras: (i) permutation invariant algebras with 
hermitean number operators and (ii) permutation invariant algebras with non-hermitean number operators. The class
(ii) has not been discussed previously and  in Subsection 3.2 several new results for those algebras were given. 
Both classes have been treated in three completely equivalent ways, first proposed in [7]. In the analysis, we have 
used concepts of a 
normally ordered expansion (18), the action of annihilation operators on the states in Fock space (19) and the  
notion of Gram matrices of scalar products in Fock space (20). We have found the general structure of the number
operators, Eqs.(23,24), and the exchange operators, Eqs.(25)-(27). The results of this section have been applied in
Section 4 to the $S_M$-extended Heisenberg algebra (29), underlying the M-body Calogero model. While the previous analyses of
this algebra were performed mainly on the symmetric ( or antisymmetric ) subspace of the whole Fock space
[6], here we  have analysed  the whole Fock space of states. 
  Our main results are the following.\\
     We have rewritten the $S_M$-extended Heisenberg algebra in the form of the  
( generalized ) triple operator algebra (31). This is a generalization of the known result for the
single-mode case (3). Then, we have found the action of annihilation operators on the monomials in Fock 
space (35). Using this, we have calculated  one- and  two-particle  Gram matrices and discussed their structure and
eigensystem. We have found that there exists a universal critical point in  Fock space, given by 
$\nu =-\frac{1}{M}$, and all states in Fock space have positive norms for $\nu > -\frac{1}{M}$. As we
have made a comment in Remark 4, the same critical point exists for a large class of  $S_M$-extended Heisenberg algebras.
Then, we have proceeded with the construction of number operators and exchange operators. We have given  explicit examples of
the structure of these operators in the case of $M=3$, Eqs.(37)-(39). Generalizing the construction of the dual
algebra from Section 2, we have defined a dual multi-mode algebra in terms of the operators
$(\tilde{a}_i,\tilde{a}_i^{\dagger})$ (40,41). With this operators, we have been able to write the number operators in
a compact form. Finally, we have shortly discussed the existence of mapping from the $S_M$-extended Heisenberg algebra (29) 
to  Bose oscillators.\\
We note that the Calogero model has been related [25] to $q$-deformed quantum mechanics [26]. The ideas presented
here may help in elucidating the connection between algebraic structures arising from the deformation of the phase
space of  ordinary quantum mechanics and Calogero-type models.

\bigskip

{\it Acknowledgements}. We would like to thank L. Jonke for useful discussions.
This work was supported by the Ministry of Science and Technology of the
Republic of Croatia under Contracts No. 00980103 and No. 0119222.

\newpage


{\bf A}$\;${\bf Appendix}\\

Here, we give two examples of complete Gram matrices for $M=2$ and $M=3$ oscillators and  
two - particle states.

{\it Example} {\it A.1} The Gram matrix for $M=2$. The matrix is written in the basis\\
\{ $a_1^{\dagger 2}|0\rangle $, $a_2^{\dagger 2}|0\rangle $, $a_1^{\dagger }a_2^{\dagger }|0\rangle $,
$a_2^{\dagger }a_1^{\dagger }|0\rangle $\}.
\begin{center}
$$
\left(\begin{array}{cccc}
a & b & b & b \\
b & a & b & b\\
b & b & d & d\\
b & b & d & d
\end{array}
\right) ,
$$
\end{center}
where $a=2+3\nu $, $b= -\nu$ and $d=1+\nu $.

{\it Example} {\it A.2} The Gram matrix for $M=3$. The matrix is written in the basis\\
 \{ $a_1^{\dagger 2}|0\rangle $, $a_2^{\dagger 2}|0\rangle $, $a_3^{\dagger 2}|0\rangle $, 
$a_1^{\dagger }a_2^{\dagger }|0\rangle $, $a_1^{\dagger }a_3^{\dagger }|0\rangle $,
 $a_2^{\dagger }a_3^{\dagger }|0\rangle $, $a_2^{\dagger }a_1^{\dagger }|0\rangle $, 
$a_3^{\dagger }a_1^{\dagger }|0\rangle $, $a_3^{\dagger }a_2^{\dagger }|0\rangle $ \}.

\begin{center}
$$
\left(\begin{array}{ccccccccc}
a & b & b & b & b & c & b & b & c\\
b & a & b & b & c & b & b & c & b\\
b & b & a & c & b & b & c & b & b\\
b & b & c & d & b & b & d & b & b\\
b & c & b & b & d & b & b & d & b\\
c & b & b & b & b & d & b & b & d \\
b & b & c & d & b & b & d & b & b\\
b & c & b & b & d & b & b & d & b\\
c & b & b & b & b & d & b & b & d
\end{array}
\right) ,
$$
\end{center}
where $a = 2 + 6\nu + 2\nu^2$, $b = - \nu - \nu^2$, $c = 2 \nu^2$ and $d = 1 + 3 \nu + 2 \nu^2$.\\
It is straightforward to write two-particle Gram matrices for any $M$. Non - zero matrix elements are of the
type ( $(i,j,k,l) = 1,2...M $ ) 
$$
\langle 0|a_i^2a_i^{\dagger 2}|0\rangle \equiv a = [1+\nu (M-1)][2+\nu (M-1)]-\nu^2 (M-1) ,
$$
$$
\langle 0|a_i^2a_j^{\dagger 2}|0\rangle = \langle 0|a_ia_ja_i^{\dagger 2}|0\rangle = 
\langle 0|a_ia_ja_j^{\dagger 2}|0\rangle = \langle 0|a_ia_ja_k^{\dagger } a_j^{\dagger }|0\rangle =
$$
$$
=\langle 0|a_ia_ja_i^{\dagger } a_k^{\dagger }|0\rangle \equiv b = -\nu -\nu^2 (M-2) ,
$$
$$
\langle 0|a_ia_ja_k^{\dagger 2}|0\rangle =\langle 0|a_ia_ja_k^{\dagger } a_l^{\dagger }|0\rangle \equiv c = 
2\nu^2 ,
$$
$$
\langle 0|a_ia_ja_i^{\dagger } a_j^{\dagger }|0\rangle \equiv d =
[1+\nu (M-1)][1+\nu (M-2)] .
$$
It is understood that different indices are not equal.

\newpage



\end{document}